Bending of core–shell nanowires by asymmetric shell deposition

*Spencer McDermott and Ryan B. Lewis**

Department of Engineering Physics, McMaster University, L8S 4L7 Hamilton, Canada
*Email: rlewis@mcmaster.ca

Abstract

Freestanding semiconductor nanowires have opened up new possibilities for semiconductor devices, enabling geometries, material combinations and strain states which were not previously possible. Along these lines, spontaneous bending in asymmetric core–shell nanowire heterostructures has recently been proposed as a means to realize previously unimagined device geometries, novel strain-gradient engineering and bottom-up device fabrication. The synthesis of these nanostructures makes use of the nanowire geometry and the directionality of the shell deposition process. Here, we explore the underlying mechanisms of this bending process by following the evolution of nanowires during asymmetric shell deposition. We show how bending can lead to dramatic local shell thickness, curvature and strain variations along the length of the nanowire, and we elucidate the dependence of shell growth and bending on parameters such as the core and shell dimensions and materials, and angle of incidence of the deposition source. In addition, deposition shadowing by neighboring nanowires is explored and employed to connect nanowire pairs, which could be used to fabricate novel nanowire sensors. GaAs–InP and GaAs–(Al,In)As core-shell nanowire growth experiments are compared to model results. These findings will guide future experiments and help pave the way to bent nanowire devices.

Introduction

The mechanical flexibility of nanowires has recently opened up new possibilities for unconventional strain engineering in semiconductor devices. For example, immense strains of up to 16%—approaching the theoretical elastic limit—have been realized in mechanically stretched Si nanowires,[1] and misfit strain in GaAs-based core–shell nanowire heterostructures has enabled the bandgap of GaAs to be adjusted by up to 40%.[2] In addition to uniform strain fields, bending in nanowires presents a unique opportunity to engineer large strain gradients and to control the nanowire geometry. Bending has been achieved by mechanical manipulation of nanowires [3–8], and spontaneously in asymmetric lattice-mismatched core–shell heterostructures. Asymmetric core–shell bent nanowires have been achieved with III-V materials[9–16] including nitrides,[17] group IV,[18,19] [20] and semiconductor–superconductor hybrids.[10,20,21] The resulting strain gradients can induce quasielectric,[4,5,13] piezoelectric,[3,22,23] and flexoelectric[23,24] fields, which can be employed to control the motion of charge carriers. Strain gradients have been used to enhance the emission intensity of quantum dots (QDs) embedded within the nanowires by more than an order of magnitude compared to straight nanowires.[13] These effects have the potential to lead to faster and more efficient optoelectronic devices. In 2021, the photoconductive properties of In(As,Sb) core–shell bent nanowires were reported.[16]

Fabrication techniques using directional deposition sources have a natural ability to produce the asymmetric nanowire heterostructures that lead to bending. We previously demonstrated that the directionality of molecular beam epitaxy (MBE) is well suited for controlled bending by asymmetric nanowire shell growth.[13] More recently, controlled bending was also demonstrated using metal–organic molecular beam epitaxy (MOMBE) using metal–organic III sources,[14] as well as electron beam evaporation of transition metals on silicon nanowires.[19] Nano x-ray diffraction analysis on highly bent nanowire heterostructures has revealed large strain and shell thickness variations along the nanowire length.[15] These results highlight the need for a more detailed investigation into how bending and asymmetric shell deposition proceeds in this novel bottom-up bent nanowire growth process.

Theoretical modeling has been used to elucidate bending in asymmetric lattice-mismatched core–shell nanowires by revealing the relationship between shell thickness, asymmetry, and composition on bending curvature.[12,14,25] Modeling has shown that thermal expansion/contraction effects have a negligible effect on bending, with bending resulting from relaxation of the lattice strain.[14] Additionally, a thermodynamic model was used to account for both surface and strain energy effects in the bending of thin nanowires.[25] While these models help to elucidate asymmetric core–shell bending, the effects of bending on the growth process itself—essential for bent nanowire fabrication—have not been explored.

The shadowing of deposition fluxes by neighboring nanowires has been shown to affect nanowire core morphology[26–28] and asymmetric shell growth.[13,19] This flux shadowing has posed challenges for the growth of dense core–shell nanowire arrays. However, the ability to locally influence the deposition on a nanowire by a neighboring object also presents opportunities.

In this work, we model the synthesis of bent nanowires by asymmetric lattice-mismatched shell deposition by a directional flux. We show how bending affects the local deposition along the nanowire, resulting in large variations in the shell thickness, local curvature and strain along the nanowire, as well as the overall nanowire shape and bending angle. The impact of core and shell dimensions and materials, as well as the geometry of the deposition system are shown to crucially influence the bending process. Finally, shadowing between nanowire pairs is modeled, presenting a promising means to further control the nanowire geometry and an easier approach for connecting nanowires to form devices. Modeling results are compared to growth experiments of asymmetric GaAs–InP and GaAs–(Al, In)As core–shell bent nanowires. The goal of this paper is to demonstrate the principles of asymmetric nanowire shell growth, and present opportunities for employing bending in nanowire device fabrication.

Methods

Bending in isolated nanowires and nanowire pairs was modeled using GNU Octave. Asymmetric core–shell nanowires with circular cores were explored and shell deposition was assumed to be perfectly directional (i.e., surface diffusion and flux divergence were neglected). A schematic of a core–shell nanowire cross section is shown in figure 1a. Nanowires are divided into 25-nm-long segments along their axis. Within a given segment, the core and shell thicknesses, and segment curvature are assumed constant. However, these properties can vary between segments and along the nanowire length. Asymmetric shell deposition is modeled as follows: nanowire cores are initially straight and standing vertically on the substrate. A directional shell deposition is incident at an angle θ with respect to the substrate normal. The deposition process is broken into steps (each step corresponding to planar deposited thickness 0.05–0.43 nm), and the local deposition on a segment varies as the cosine of the average angle between the segment normal (perpendicular to the nanowire axis) and the incident flux. After each deposition step, the curvature of each segment is calculated from linear elastic theory,[13] and the segments are linked to create the complete bent nanowire. The resulting nanowire geometry is used to carry out the next deposition step, and this process is repeated until the shell deposition is completed. There is no deposition on a nanowire segment if the angle between the segment normal and the incident flux is >90°. In this case, the deposition on the nanowire is self-shadowed by the nanowire itself. Self-shadowing is illustrated in Figure 2a. Similarly, a neighbouring nanowire (or other object) could shadow the flux. This is determined by projecting the flux from the tip of the first nanowire (or object) to the shadowed nanowire. If the segment is below the projected path it is shadowed. Shadowing between nanowire pairs is illustrated in Figure 4b. In both cases we model shadowing to be fully effective (no deposition on shadowed areas).

Shells are assumed to be coherently strained and have a positive lattice mismatch from the core. The accumulation of positive-lattice-mismatched material asymmetrically deposited on the nanowire results in compressive axial strain in the shell and tensile axial strain in the core. This strain is calculated by as described in the Supporting Information of Lewis *et al.*[13] :

Assuming the nanowire is constrained to be straight, the core and shell share a single axial lattice constant, $a_{\text{interface}}$, given by equation 1. In this case, we define the resulting strain in the core and shell with equations 2 and 3.

$$a_{\text{interface}} = \frac{a_{core} a_{shell}(a_{core} A_{shell} E_{she} + a_{shell} A_{core} E_{core})}{a_{core}^2 A_{shell} E_{shell} + a_{shell}^2 A_{core} E_{core}} \quad (1)$$

$$\epsilon_{core} = \frac{a_{\text{interface}} - a_{core}}{a_{core}} \quad (2)$$

$$\epsilon_{shell} = \frac{a_{\text{interface}} - a_{shell}}{a_{shell}} \quad (3)$$

Analogous to a bimetallic strip, strain in an unconstrained wire can be partially relieved by bending—resulting in a linear strain gradient along a bisecting line (equation 4), such as that shown in figure 1a.

$$\epsilon = \begin{cases} \epsilon_{core} + \kappa(x - \bar{x}_c) & x < x_{interface} \\ \epsilon_{shell} + \kappa(x - \bar{x}_c) & x \geq x_{interface} \end{cases} \quad (4)$$

Where $\kappa$ is the curvature of the nanowire and $x$ is the position along the bisecting line from right to left in figure 1a. The bending strain gradient is directly proportional to the nanowire's curvature. The strain from bending reduces the total strain energy in the nanowire compared to the straight case. The point where there is zero bending strain ( $\bar{x}_c$ ) is the Young's Modulus weighted centroid. The Young's Modulus weighted centroid is the average centroid of the core and shell weighted by their Young's Moduli and cross-sectional area (equation 5)

$$\bar{x}_c = \frac{\sum_{i=1}^{N} A_i E_i \bar{x}_i}{\sum_{i=1}^{N} A_i E_i} \quad (5)$$

Where $A_i$ is the cross-sectional area of component $i$, $E_i$ is the Young's modulus, $\bar{x}_i$ is the centroid of the component along the bisection, and $N$ is the total number of components of the nanowire [core plus shell(s)].

The curvature of the nanowire segment is found by minimizing the total elastic potential with respect to the segment curvature. Points are generated along the bisection to numerically compute the elastic potential. The trapezoidal method gives the elastic potential over the cross section (equation 6).

$$U = \int_{x_{shell}}^{x_{core}} w(x) E(x) \epsilon(x)^2 \delta x \quad (6)$$

Where $E(x)$ is Young's modulus, $\epsilon(x)$ is strain, and $w(x)$ is the width along the bisection. $x_{core}$ and $x_{shell}$ are the boundary points on the bisection—the innermost surface of the core and the outermost surface of the shell.

Results and discussion

Figure 1b shows the local segment curvature as a function of shell thickness for GaAs–InP core–shell nanowires with three different core diameters ($D$). The maximum curvature occurs for a shell thickness of 0.42× the core diameter. The maximum curvature is inversely proportional to diameter and obeys equation 7.

$$D_1 \cdot \kappa_{max}(D1) = D_2 \cdot \kappa_{max}(D2) \quad (7)$$

Where $D1$ and $D2$ represent two independent nanowire diameters with the same core–shell material. Therefore, the product of maximum curvature and diameter is a constant which depends on the materials composing the nanowire core and shell $(D_1 \cdot \kappa_{1,\,max})$. For GaAs–InP nanowires, this constant is 0.031.

Results and Discussion

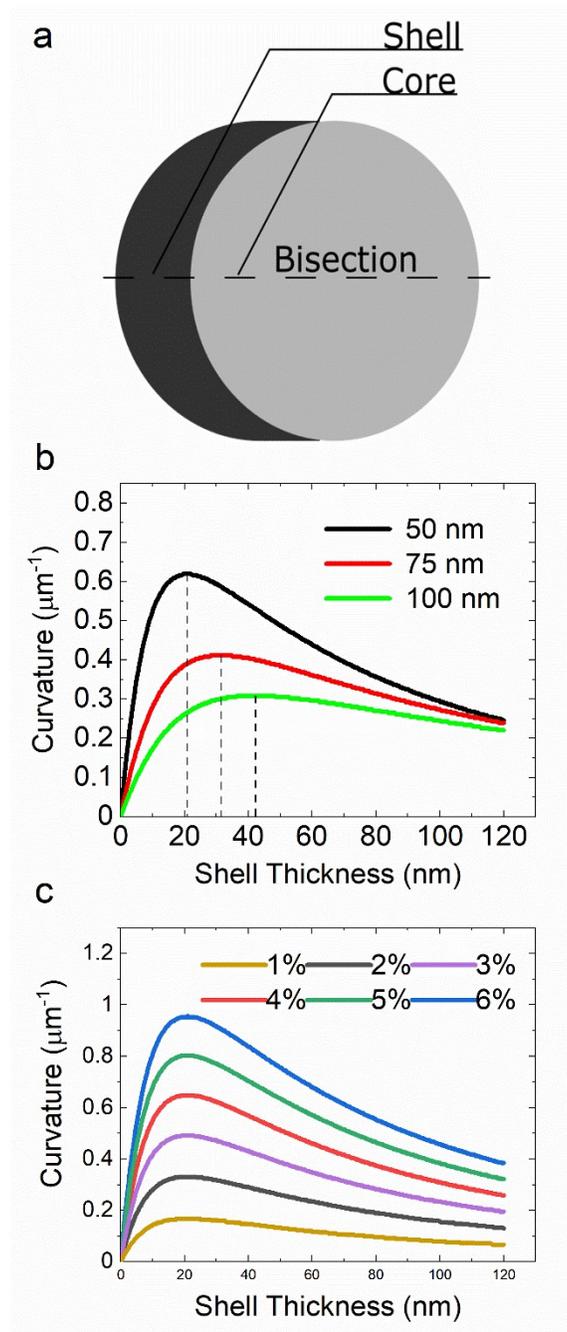

Figure 1. (a) Schematic representation of the nanowire core–shell cross section geometry with a dotted bisection line. (b) Nanowire segment curvature as a function of shell thickness for three GaAs–InP core–shell nanowires (lattice mismatch 3.8%) with core diameters: 50 nm, 75 nm, and 100 nm. The maximum curvature occurs for a shell thickness of 0.42× the nanowire diameter, as indicated with vertical dotted lines in (b). (c) Nanowire curvature as a function of shell thickness for different core–shell lattice mismatches for a 50 nm diameter core.

In Figure 1b the curvature peak broadens with increasing core diameter. The full width at half maximum (FWHM) is 93 nm for the 50 nm diameter core and scales linearly with diameter. For a fixed core–shell thickness ratio, the curvature is inversely proportional to the core diameter. For a nanowire with uniform curvature, the bending angle—the angle between the substrate normal and the axis of the tip of the nanowire—is given by the product of the nanowire length and curvature. Consequently, the maximum achievable bending angle is determined by the nanowire core aspect ratio. We note that for real nanowires, the relation between core diameter and length depends nontrivially on the growth process.

Figure 1c presents the local curvature as a function of shell thickness for nanowire segments with 50 nm diameter cores and various core–shell lattice mismatches. The curvature is directly proportional to lattice mismatch, so all curves have the same shape and the peak curvature occurs at a shell thickness of 42% of the core diameter. The core and shell Young's moduli were taken as those of GaAs (85.5 Gpa) and InP (61.1 GPa), respectively. There is a small effect from Young's modulus differences between the core and shell: increasing the shell modulus increases and sharpens the peak curvature, and the peak occurs with a thinner shell.

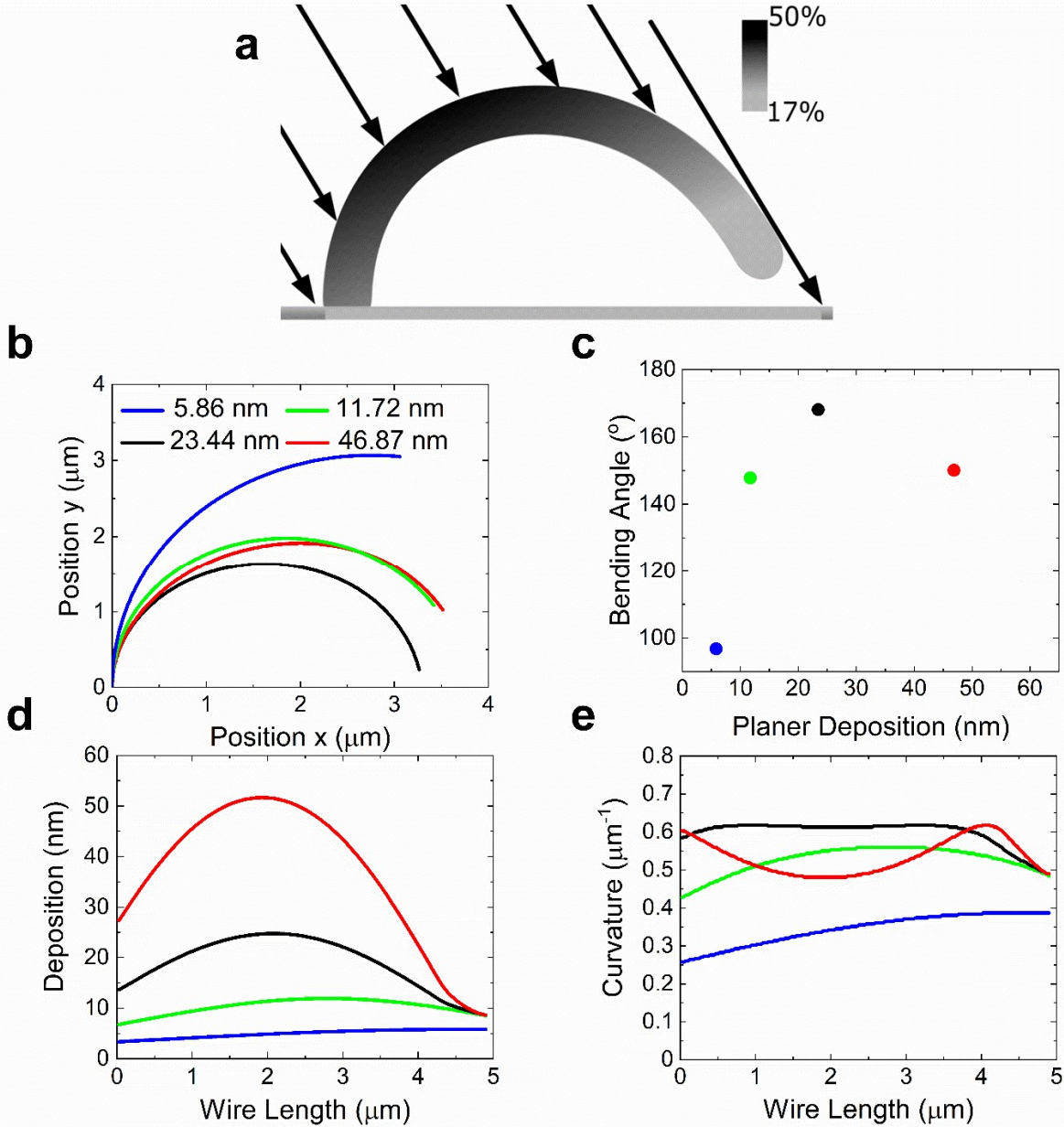

Figure 2. (a) Schematic representation of a deposition flux (indicated by black arrows) incident on an optimally bent GaAs–InP nanowire. The nanowire shading indicates the local shell thickness relative to core diameter. (b, c, d, e) modeling GaAs–InP core–shell nanowires with 50 nm core diameter and aspect ratio of 85 for four planer shell depositions: 5.9, 11.7, 23.4, 46.9 nm at a deposition angle of 30 degrees. (b) The wire geometries projected on a 2d plane. (c) The bending angle of the nanowire tips plotted as a function of planer deposition. (d) The shell thickness along the wire length from base to tip. (e) The curvature along the wires. The legend in (b) corresponds to all panels.

During shell growth, the local instantaneous shell deposition rate depends on the local orientation of the nanowire segment relative to the incident flux. Consequently, while the deposition rate is uniform along the length of a straight nanowire, the deposition rate will vary along the length of a bent nanowire and evolve throughout the deposition as the nanowire bends. Figure 2a illustrates a bent nanowire under an incident flux. The effective flux on the nanowire sidewall varies along the length of the bent nanowire as a result of the changing local angle between the nanowire surface normal and the incident flux. Variation in total local shell thickness is indicated by shading of the nanowire, showing the shell to be thinner at the base and tip compared to the middle, for this geometry. The tip of this nanowire is self-shadowed.

The shape and bending angle of GaAs–InP nanowires with a core diameter of 50 nm and an aspect ratio of 85 are shown in Figures 2b,c for different stages of shell growth corresponding to four planar deposition thicknesses. Planar depositions of 5.9 and 11.7 nm correspond to fully under deposited shells (local shell thickness below that corresponding to maximum curvature along the entire nanowire length). In this case, further shell growth will increase the curvature everywhere along the nanowire. The 23.4 nm thickness corresponds to near optimal deposition, resulting in the maximum possible bending angle and average nanowire curvature. Finally, the 46.9 nm thickness results in an over-deposited shell. In this case, the bending angle and average curvature along the nanowire is lower than for the optimally-bent nanowire, and further deposition will further reduce the bending angle.

Figure 2d,e show the resulting shell thickness and local curvature along the length of the nanowires for the deposition thicknesses presented in figure 2b. With increasing shell growth, the deposition shifts from the tip of the nanowire towards the nanowire base. The 23.4 nm planar deposition corresponds to an average shell thickness of 19.0 nm, close to the optimal of 22.5 nm (42% of the core dimension). However, we note that in this case the local thickness varies by a factor of two along the nanowire. Figure 2e illustrates the complex dependence of the local curvature on the distance along the nanowire and the deposition thickness. Initially, when the shell is fully under-deposited, the local curvature follows the same trend as the local deposition thickness, showing an initial peak at the nanowire tip which shifts towards the base with increasing shell growth. At optimal shell deposition, the middle of the nanowire is slightly over-deposited, while the tip and base are still under-deposited. Under these conditions, the curvature is remarkably constant along a large portion of the nanowire (a result of the slowly decreasing curvature with over-deposition). For the largest shell deposition displayed, the curvature is now peaked at the base and near the tip, with the highly over-deposited middle showing less curvature. These results illustrate the complex evolution of the nanowire shape, shell thickness and curvature profile along the nanowire with deposition.

The local deposition rate follows the cosine of the angle between the flux and the segment normal: exhibiting a maximum when the segment is perpendicular to the flux. However, when this angle exceeds 90 degrees the nanowire surface is self-shadowed and local deposition does not occur. Figure 2a illustrates self-shadowing at the nanowire tip. Significant self-shadowing has also occurred for the two largest deposition thicknesses in figures 2d–e, explaining why the shell thickness at the nanowire tip remains largely unchanged above 11.7 nm planar deposition.

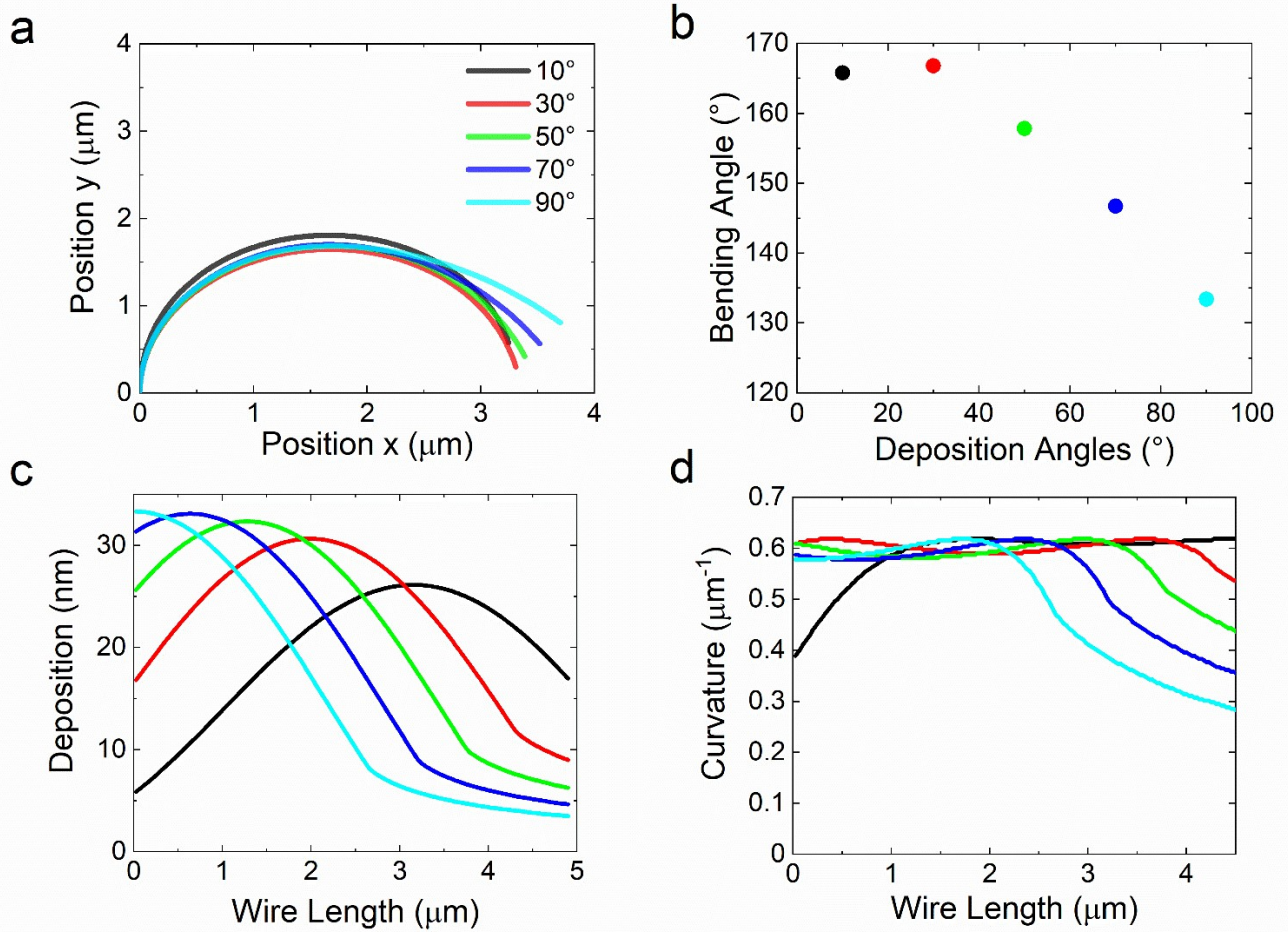

Figure 3. Effect of varying the angle of the incident flux with respect to the substrate normal for GaAs–InP wires. A deposition of 33.3 nm in the direction parallel to the incident flux was considered for flux angles of 10, 30, 50, 70, and 90 (corresponding to 32.8, 28.9, 21.4, 11.4, and 0 nm plainer depositions, respectively), as indicated in the legend in (a), which corresponds to all panels. The wires are 4 μm long with a 50 nm core diameter. (a) Wire geometry projected on a 2d plane. (b) Tip bending angle plotted as a function of incident flux angle. (c) Local shell thickness along the wire length. The decreasing thickness toward the nanowire tip with increasing deposition angle results from self-shadowing. (d) Local curvature along the wire length.

The deposition angle of the incident flux (fixed by the growth system configuration) has a significant impact on the shell deposition profile and the resulting nanowire geometry. Figure 3a,b present the nanowire side profile and the bending angle of the nanowire tip, respectively, for shells deposited with varying incident flux angles on nanowire cores of 4 μm length and 50 nm diameter. Figure 3a reveals that the final shape is considerably affected by the flux angle—particularly at the tip of the nanowire. At the beginning of shell deposition—nanowire standing vertically on the substrate—the 90 degree flux angle results in the highest shell growth rate. However, this situation changes as the nanowire bends throughout the deposition. As shown in Figure 3b, the highest bending angle is achieved for a flux angle of 30 degrees for these conditions. Notably, the amount of deposition affects this result—if deposition continued the 10 degree flux could create a higher bending angle than the 30 degree flux as less shadowing occurs at the nanowire tip. The low initial growth rate for the 10 degree flux causes less deposition overall, resulting in a lower bending angle than for the 30 degree flux. Figure 3c reveals how the shell deposition is distributed along the nanowire for different flux angles. For higher deposition angles, the deposition is shifted towards the base of the nanowire, peaking at the base for 90 degree deposition. Furthermore, the maximum local deposition increases with increasing deposition angle. In contrast, the onset of self-shadowing occurs sooner for higher flux angles, resulting in a larger shadowed region towards the tip of the nanowire, as indicated in figure 3c by the reduced shell thickness toward the nanowire tips. This shadowing reduces the tip curvature, as seen in the curvature plot displayed in figure 3d. These results show that the geometry, bending angle, shell thickness and curvature along the nanowire depend nontrivially on the deposition angle, as well as the other growth conditions. While the exact behaviour depends on other deposition conditions (e.g., nanowire geometry, deposition amount), in general, shell deposition shifts towards the base of the nanowire as the flux angle increases. Depending on the other deposition conditions, this can have different implications for the bending angle. The large impact of flux angle on shell growth is in contrast to conventional planar layer growth, where the flux angle is essentially irrelevant.

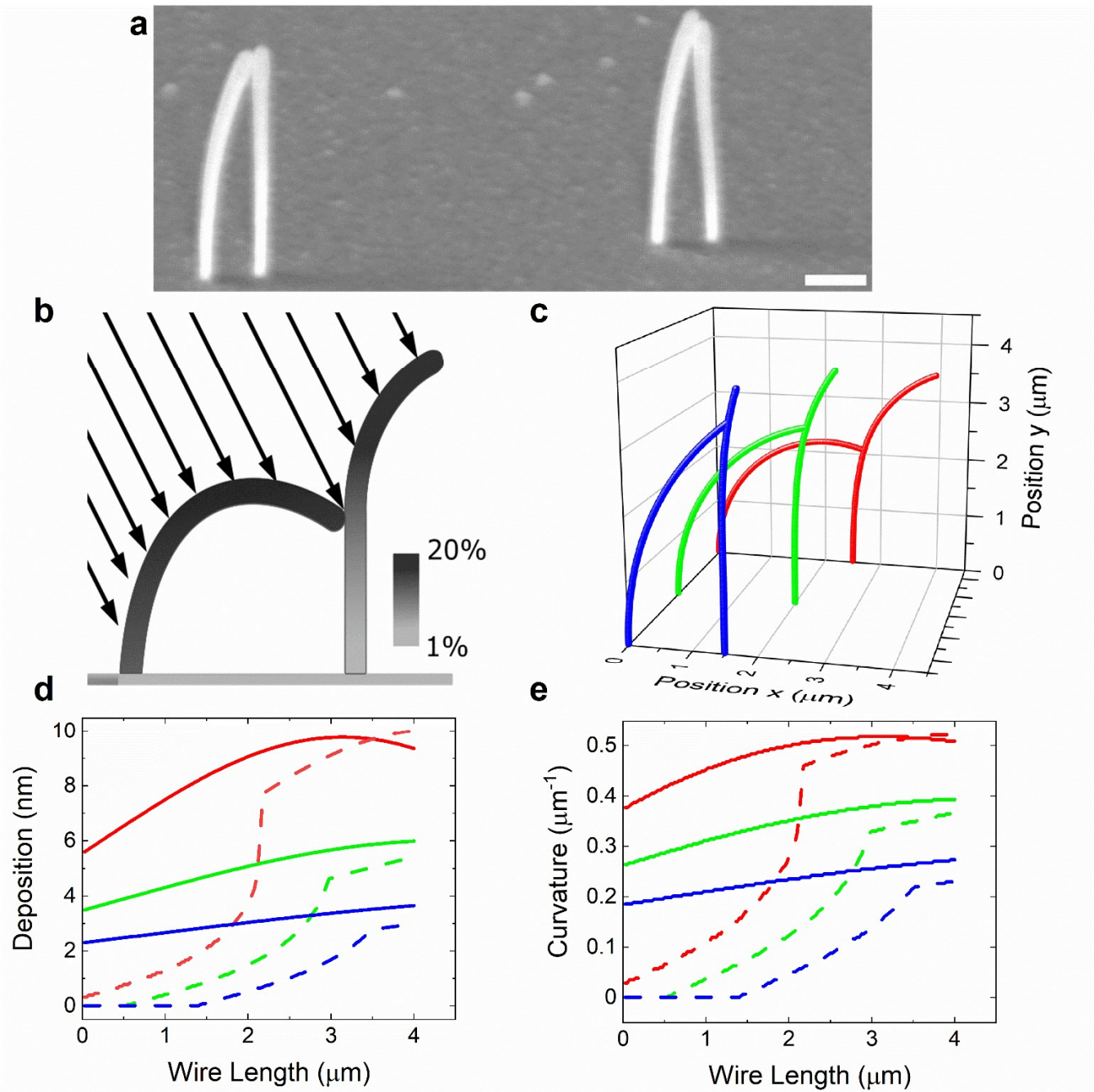

Figure 4. (a) Scanning electron micrograph of GaAs–InP core–shell nanowire pairs, separated by 600 nm and bent to the point of contact as a result of pair shadowing; the sample is tilted 30°, and the scale bar is 1 μm. (b) Schematic illustration of one nanowire shadowing the incident flux from another; the scale bar represents shell thickness relative to core diameter. (c) 3D plot of modeled wire geometry of three GaAs–InP nanowire pairs with different separation distances, bent by deposition incident from the left until the point of contact. The pair separations are (blue) 1.5, (green) 2, (red) 2.5 μm and the corresponding planer deposition are 4.0, 6.0, 9.7 nm, respectively. The deposition angle is 30 degrees, and the core aspect ratio is 80 with a diameter of 50 nm. (d)Deposition and (e) curvature along the wire

length for the pairs shown in (c). The solid and dashed lines correspond to the unshadow and shadowed nanowires, respectively.

In addition to self-shadowing, the flux incident on a nanowire can also be shadowed by neighboring nanowires. Figure 4a presents a scanning electron micrograph (SEM) of pairs of MBE-grown asymmetric GaAs–InP core–shell nanowires bent to the point of contact. These wires are of a similar composition to Wallentin et al.'s [11] GaAs–GaInP bent core–shell nanowires. As discussed in our previous report, [13] contacting of nanowire pairs with uniform curvature would require bending angles of $\geq 90$ degrees. Clearly, the nanowires in Figure 4a are bent by much less. In this case, contacting is a result of the different bending profiles of the two nanowires—a result of one nanowire shadowing the flux incident on the other. We modeled pair shadowing assuming that a nanowire will fully block the flux on a portion of a shadowed nanowire if it obstructs the path between that portion of the nanowire and the flux. This assumes a non-divergent flux source and that the nanowire pairs and the incident flux reside in a single plane. Deviations in the alignment of the nanowires as well as broadening of the incident flux angle distribution in the azimuthal direction would result in a reduction in the shadowing effect. Thus, these results represent an upper limit on the effect of pair flux shadowing. Figure 4b schematically illustrates pair shadowing, where the flux is partially obstructed from depositing on the nanowire on the right by the nanowire on the left. Figure 4c presents model results of three different GaAs–InP nanowire pairs with separations of 1.5–2.5 µm, deposited on until the nanowires are just touching. The cores are 4 µm long and 50 nm in diameter and the flux is incident at 30 degrees from the substrate normal. The leading nanowire contacts the shadowed nanowire when it bends a horizontal distance approximately equal to the initial separation, due to the negligible bending in the lower portion of the shadowed nanowire. Pair shadowing can dramatically reduce the amount of bending required to connect nanowire pairs, especially if the nanowires are in close proximity. Figure 4d–e shows the deposition and curvature along nanowire pairs. The impact of shadowing increases when nanowires are placed closer together. Pair shadowing initially only affects the lower portion of a shadowed nanowire, however, this impacts the bending which in turn impacts the deposition profile even at the unshadowed tip portion of the nanowire. These results demonstrate how flux shadowing can be employed to further control the shape of bent nanowires, in particular to connect nanowires together for device fabrication. Bent nanowire devices have already been fabricated E.g., Chemical field-effect transistors (chem-FET) by mechanical bending. [6] Pair shadowing presents an easier path for using bending to achieving multiple electrical/optical connections to nanowire-based devices, such as bio sensors.

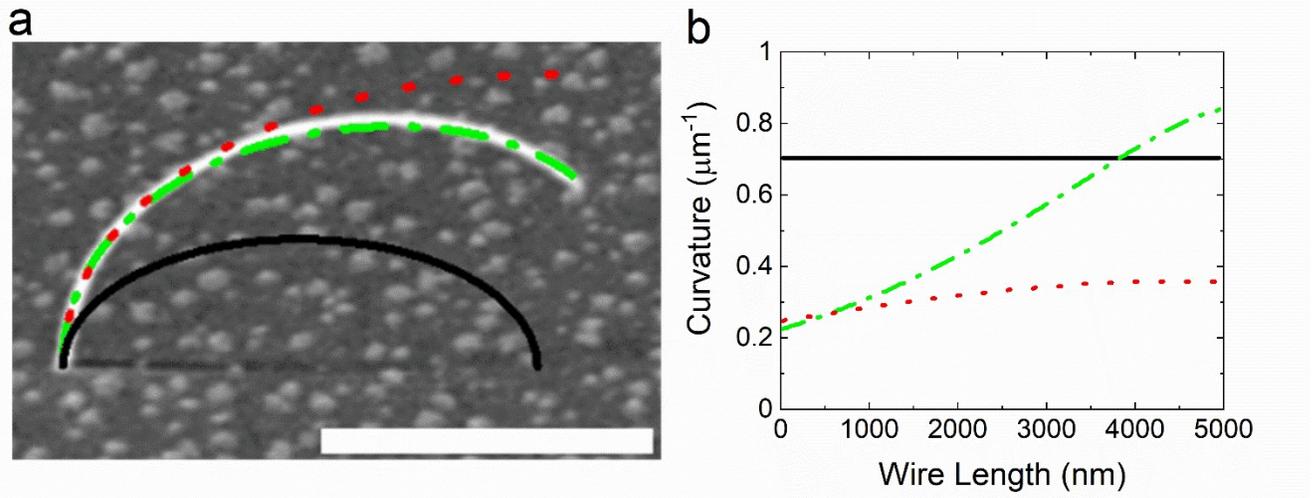

Figure 5. (a) SEM image of an asymmetric GaAs–Al0.5In0.5As core–shell nanowire pair from[13] and fit with various shell deposition models. Four model fits are shown: (black line) uniform, unshadowed deposition with the nominal parameters from[13] (uniform 20 nm shell and 45 nm core diameter); (red dots) core diameter of 45 nm with a 3.9 nm planner deposition according to our model; (green dot–dashes) linear variation in core diameter from 61 nm at the base to 30 nm at the tip and a planner deposition of 6.9 nm The plots have been rescaled to account for the imaging angle of 45° from the substrate plane. The scale bar is 2 μm for the SEM image. (b) Curvature along the wire length from the three models.

To further validate our bending model, we compared it to an asymmetric GaAs–Al0.5In0.5As core–shell nanowire from.[13] An SEM image of the nanowire is shown in Figure 5a along with modeling results for four sets of conditions: first, (black curve) the nominal conditions of a uniform shell thickness of 20 nm deposited on a uniform core of 45 nm thickness. These conditions greatly overestimate the curvature of the nanowire. Second, (red curve) 4.7 nm of deposition on a uniform 45 nm core following our model. This results in a much better fit, however it underestimates the curvature towards the tip of the nanowire. Third, (green curve) a planner deposition of 6.9 nm according to our model with a linearly varying core diameter from 61 to 30 nm base to tip. This core thickness variation is consistent with SEM images of the nanowire cores in the supporting information of Lewis et al.[13] Also, the results are consistent with previous modeling done by Greenberg et al.[14] where it is demonstrated that excluding tapering in the nanowire leads to less bending in the model then the experimental result. Figure 5b depicts the curvature along the nanowire for the three models described above. Uniform deposition causes a uniform curvature as seen by the first model (black line). The model corresponding to the tapered core exhibits a rapidly increasing curvature from base to tip, consistent with the actual shape of the nanowire in the SEM. These results show how the shape of this wire results from the complex shell deposition process modeled here. Furthermore, they show that tapering in the nanowire core can further complicate the final nanowire shape.

Conclusion

A model was developed to explore the underlying mechanisms in nanowire bending from asymmetric shell deposition. Nanowire curvature is limited by core diameter and core–shell lattice mismatch, while the nanowire bending angle is limited by the aspect ratio of the core. Shell deposition and local curvature varies along the length of a nanowire—a consequence of the evolution of the projected flux on the nanowire sidewall, which varies throughout the deposition and along the nanowire as it bends. Our results demonstrate that these effects can have dramatic consequences for the overall growth and nanowire geometry. For instance, in highly bent nanowires, self-shadowing of the nanowire by itself can block deposition completely. Shell deposition is strongly dependent on the angle of the incident flux from the substrate normal, in contrast to conventional planar layer growth. Shadowing in nanowire pairs was modeled and presents an easy path to connect nanowires without the need for high amounts of bending. Such connected nanowire structures are prospective for nanowire sensors or for electrical/optical coupling between elements on a chip. Our modeling results are consistent with experimental observations from asymmetric GaAs–InP and GaAs–(Al,In)As core–shell nanowires. These findings present new considerations and opportunities for controlling the geometry and strain in nanowires and for enabling the bottom-up fabrication of connected nanowire devices.

Acknowledgments:

We gratefully acknowledge financial support from the Natural Sciences and Engineering Research Council of Canada [RGPIN-2020-05721]. The authors are also grateful for assistance from the Centre for Emerging Device Technologies and the Canadian Centre for Electron Microscopy at McMaster University, and the Toronto Nanofabrication Centre at the University of Toronto.

Lastly, we would like to thank John W. Eaton, David Bateman, Søren Hauberg, and Rik Wehbring for there work on GNU Octave version 5.2.0 manual: a high-level interactive language for numerical computations.